# Decoding and mapping task states of the human brain via deep learning

Xiaoxiao Wang[1#], Xiao Liang[1#], Zhoufan Jiang[1], Benedictor Alexander Nguchu[1], Yawen Zhou[1], Yanming Wang[1], Huijuan Wang[1], Yu Li[1], Yuying Zhu[1], Feng Wu[1], Jia-Hong Gao[2,3*], Benching Qiu[1*]

1 Centers for Biomedical Engineering, University of Science and Technology of China, Hefei, China.
2 MRI Research Center and Beijing City Key Lab for Medical Physics and Engineering, Peking University, Beijing, China
3 McGovern Institute for Brain Research, Peking University, Beijing, China

# These authors contributed equally to this work.

**\* Correspondence:**
Benching Qiu
bqiu@ustc.edu.cn
Jia-Hong Gao
jgao@pku.edu.cn

## Acknowledgments

This study was supported by the National Natural Science Foundation of China (91432301, 81701665, 81790650, 81790651, 81430037, 81727808 and 31421003), the Fundamental Research Funds for the Central Universities of China (WK2070000033) and the Beijing Municipal Science & Technology Commission (Z171100000117012).

## Abstract

Support vector machine (SVM) based multivariate pattern analysis (MVPA) has delivered promising performance in decoding specific task states based on functional magnetic resonance imaging (fMRI) of the human brain. Conventionally, the SVM-MVPA requires careful feature selection/extraction according to expert knowledge. In this study, we propose a deep neural network (DNN) for directly decoding multiple brain task states from fMRI signals of the brain without any burden for feature handcrafts. We trained and tested the DNN classifier using task fMRI data from the Human Connectome Project's S1200 dataset (N=1034). In tests to verify its performance, the proposed classification method identified seven tasks with an average accuracy of 93.7%. We also showed the general applicability of the DNN for transfer learning to small datasets (N=43), a situation encountered in typical neuroscience research. The proposed method achieved an average accuracy of 89.0% and 94.7% on a working memory task and a motor classification task, respectively, higher than the accuracy of 69.2% and 68.6% obtained by the SVM-MVPA. A network visualization analysis showed that the DNN automatically detected features from areas of the brain related to each task. Without incurring the burden of handcrafting the features, the proposed deep decoding method can classify brain task states highly accurately, and is a powerful tool for fMRI researchers.

**Abbreviations**

2D – two dimensional
3D – three dimensional
4D – four dimensional
ABIDE – Autism Brain Imaging Data Exchange
ADNI – Alzheimer's Disease Neuroimaging Initiative
BA – Brodmann area
BN – batch normalization
BOLD – blood-oxygenation-level dependent
CNN – convolutional neural network
COPE – contrast of parameter estimate
DBN – deep belief network
DNN – deep neural network
fMRI – functional magnetic resonance imaging
FN – false negative
FP – false positive
GLM – general linear model
HCP – Human Connectome Project
HRF – hemodynamic response function
LSTM – long short-term memory
M1 – primary motor cortex
M2 – secondary motor cortex
MNI – Montreal Neurological Institute
MVPA – multi-voxel pattern analysis
RBM – restricted Boltzmann machine
ReLU – rectified linear unit
RNN – recurrent neural network
ROC – receiver operating characteristic
ROI – region of interest
S1 – primary somatosensory area
SGD – stochastic gradient descent
SVM – support vector machine
TN – true negative
TP – true positive
WM – working memory

# 1    Introduction

For years, researchers have been attempting to decode and identify functions of the human brain based on functional brain imaging data (Dehaene et al., 1998; Haynes & Rees, 2006; Jang, Plis, Calhoun, & Lee, 2017; Poldrack, Halchenko, & Hanson, 2009; Rubin et al., 2017). The most popular among these brain-decoding methods is the support vector machine (SVM) based multi-voxel pattern analysis (MVPA), a supervised technology that incorporates information from multiple variables at the same time (B. Kim & Oertzen, 2018; Nikolaus Kriegeskorte & Bandettini, 2007; N. Kriegeskorte, Goebel, & Bandettini, 2006; Norman, Polyn, Detre, & Haxby, 2006). Despite its popularity, the SVM struggles to perform well on high-dimensional raw data, and requires the expert use of design techniques for feature selection/extraction (LeCun, Bengio, & Hinton, 2015; Vieira, Pinaya, & Mechelli, 2017). Thus, we explore in this study an open-ended brain decoder that uses whole-brain neuroimaging data on humans.

In recent years, the deep neural network (DNN), a series of model-free machine learning methods, has performed well in abstracting representations of high-dimensional data (LeCun et al., 2015). The hierarchical structure of a DNN with a nonlinear activation function enables the learning of a more complex output function than those that can be learned using traditional machine learning methods, and one that can be trained end to end. DNNs have already yielded remarkable results in medical image analyses (Cichy & Kaiser, 2019; Shen, Wu, & Suk, 2017; Vieira et al., 2017). Considering these characteristics, a DNN classifier may be suited for classifying brain states directly from a massive whole-brain fMRI time series without requiring feature selection.

Deep learning methods are effective if massive amounts of data are available for training. However, under controlled conditions, most typical neuroimaging studies have collected data from only tens to hundreds of subjects, with the purpose of identifying minor differences between different states (Horikawa & Kamitani, 2017) or groups thereof (Vieira et al., 2017). An applicable brain decoder is supposed to be able to identify these differences even with a limited amount of data. Transfer learning is widely used for training DNNs with limited medical data (Sharif Razavian, Azizpour, Sullivan, & Carlsson, 2014). It takes advantage of similar data within big datasets (Ciompi et al., 2015; Kermany et al., 2018; Wen, Shi, Chen, & Liu, 2018). Recent large fMRI projects, such as the Human Connectome Project (HCP) (Van Essen et al., 2013) and BioBank (Miller et al., 2016), allow us to access massive amounts of fMRI data. It is therefore now possible to directly train a DNN decoder by means of big fMRI data and generalize the DNN decoder for common fMRI studies.

In this study, we propose a DNN classifier that effectively decodes and maps an individual's ongoing brain task state by reading 4D fMRI signals related to the task. We illustrate the generalizability of this DNN for typical neuroimaging studies by testing the decoder on the classification of task sub-types.

# 2    Methods

## 2.1    HCP datasets

The HCP S1200 minimally preprocessed 3T data release, which contains imaging and behavioral data from a large population of young healthy adults (Van Essen et al., 2013), was used in this study. We employed data of 1,034 participants of the HCP who had performed seven tasks: emotion, gambling, language, motor, relational, social, and working memory (WM). Further details of the recruitment process, imaging data acquisition, behavior collection, and MRI preprocessing can be found in previous papers (Barch et al., 2013; Van Essen et al., 2013; Van Essen et al., 2012).

## 2.2 Preparation of fMRI time series for deep learning

We analyzed the HCP volume-based preprocessed fMRI data, which had already been normalized to the Montreal Neurological Institute's (MNI) 152 space. Most of the seven tasks were constituted by control conditions (e.g., 0-back places in the WM task and shape stimuli in the emotion task) and task conditions (e.g., 2-back in the WM task and fear stimuli in the emotion task). In each task, only one condition was selected for the next step. For tasks (emotion, language, gambling, social, and relational tasks) with only two conditions, the condition that showed a greater association with the task had priority over the other. WM and motor tasks contained more than one task condition, and we randomly chose one (2-back body for WM and right hand for motor) from the list (Table 1).

| Task | Candidate Conditions | Selected Condition | Duration of the Block (seconds) |
|------|---------------------|--------------------|---------------------------------|
| **Emotion** | Fear, shape | Fear | 18 |
| **Gambling** | Reward, loss | Loss | 28 |
| **Language** | Story, math | Present story | 20 |
| **Motor** | Right hand, left hand, right foot, left loot, tongue | Right hand | 12 |
| **Relational** | Relational, match | Relational | 16 |
| **Social** | Mental, random | Mental | 23 |
| **Working Memory (WM)** | 2-back places, 0-back places, 2-back body, 0-back body, 2-back tools, 0-back tools, 2-back faces, 0-back faces | 2-back places | 27.5 |

**Table 1. Details of the selected BOLD time series for each task.**

For each task, an input sample was a continuous BOLD series that covered the entire block and eight seconds past the block, including the post-signal of the hemodynamic response function (HRF). Furthermore, each BOLD volume was cropped from $91 \times 109 \times 91$ to $75 \times 93 \times 81$ to exclude the area that was not part of the brain. Thus, the input data varied from $27 \times 75 \times 93 \times 81$ to $50 \times 75 \times 93 \times 81$ (time$\times$x$\times$y$\times$z, TR=0.72 s). A total of 34,938 fMRI 4D data items were obtained across all tasks and subjects.

## 2.3 The DNN

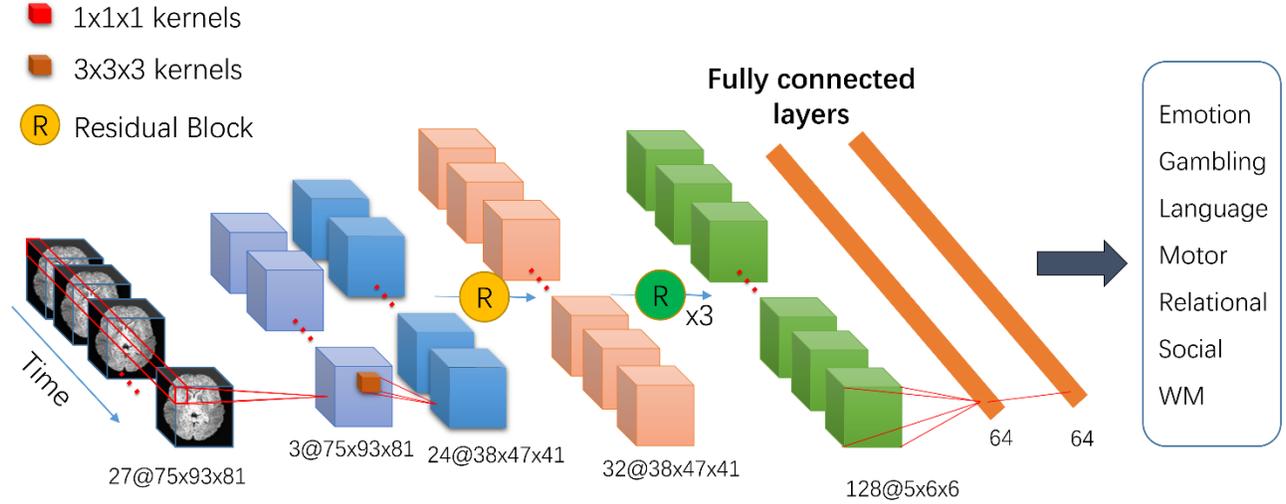

**Figure 1. The proposed deep neural network.** The network consists of five convolutional layers and two fully connected layers. The model takes fMRI scans as input and provides labeled task classes as output.

Figure 1 shows a flow diagram of our proposed network that consists of five convolutional layers and two fully connected layers. In this experiment, 27×75×93×81 data were generated via the aforementioned preprocessing and data augmentation steps. In the first layer, we used 1×1×1 convolutional filters, which have been widely used in recent structural designs of convolutional neural networks (CNNs) because these filters increase non-linearity without changing the receptive fields of the convolutional layer (Hu, Shen, & Sun, 2017; Iandola et al., 2016; Simonyan & Zisserman, 2014). These filters can generate temporal descriptors for each voxel of the volume of the fMRI, and their weights can be easily learnt by DNNs during training. Therefore, after adopting this type of filter, the time dimension of the data was reduced from 27 to three. Following this, a convolutional layer and four residual blocks were stacked to extract the high-level features. Our residual block is formed by replacing the 2D convolutional layer in the original residual block (K. He, Zhang, Ren, & Sun, 2016) with a 3D convolutional layer (Maturana & Scherer, 2015). The output channels of the four residual blocks are in *multiples of two*—32, 64, 64, and 128, respectively. We adopted a stride of two in the second convolutional layer and the last three residual blocks. These layers were designed in such a way that their dimensions could be quickly reduced to balance the consumption of GPU memory. For ease of network visualization analysis, we used a full convolution in the last convolutional layer instead of the pooling operation in CNNs used in common. Two fully connected layers were used after a stack of convolutional layers; the first had 64 channels and the second performed seven-way classification (one for each class). In our models, the rectified linear unit (ReLU) function (Krizhevsky, Sutskever, & Hinton, 2012) and batch normalization (BN) layer (Ioffe & Szegedy, 2015) were applied after each convolutional layer, whereas the softmax function was employed in the last fully connected layer.

Big data played an important role in training the DNNs. Despite the remarkable success of DNNs, their application to a limited amount of data is still a problem. Data augmentation is an efficient way to generate more samples, and has been widely used in applications (Ciompi et al., 2015; Donahue et al., 2014; Wachinger, Reuter, & Klein, 2018). The main purpose of data augmentation is to increase variations in the data where this can prevent overfitting and improve the invariance of the neural network. Contrary to traditional images, the input images in this experiment were already aligned with the standard MNI152 template; therefore, performing data augmentation in the spatial domain was considered redundant.

Considering the varied durations of the input data, we applied data augmentation in the temporal domain to improve the generalizability of the neural networks in this situation. A fragment of $k$ continuous TRs ($k$=27 in our experiments) was randomly split from each input data item in every epoch of the training stage (Figure 2a). To avoid fluctuations in the reported accuracy, only the fragment consisting of the first $k$ TRs of each data was used in validation and testing stages.

The implementation of our proposed network was based on the PyTorch framework (https://github.com/pytorch/pytorch). The design was constructed from scratch but initially utilized weights suggested by K. M. He, Zhang, Ren, and Sun (2015). To guarantee effectiveness, we used Adam with the standard parameters ($\beta_1$=0.9 and $\beta_2$=0.999) (Kingma & Ba, 2014). Due to memory constraints on the graphics board, the batch size was set to 32. The initial learning rate was set to 0.001, and gradually decayed by a factor of 10 each time the validation loss plateaued after 15 epochs. To avoid overfitting, we used the early stopping approach, and stopped training when the validation loss reached a minimum.

Our validation strategy employed a five-fold cross-validation across subjects. Prior to training, the subjects' data were categorized into subsets as follows: training set (70%), validating set (10%), and testing set (20%) (Figure 2a). The sample of training/validation/testing was later altered for each of five folds. Applying the SVM-MVPA to tens of thousands of data items is time consuming. A comparison between the SVM-MVPA and the proposed method was thus not applied to the entire dataset, but to the Test-Retest task-fMRI group data in the Transfer Learning Section.

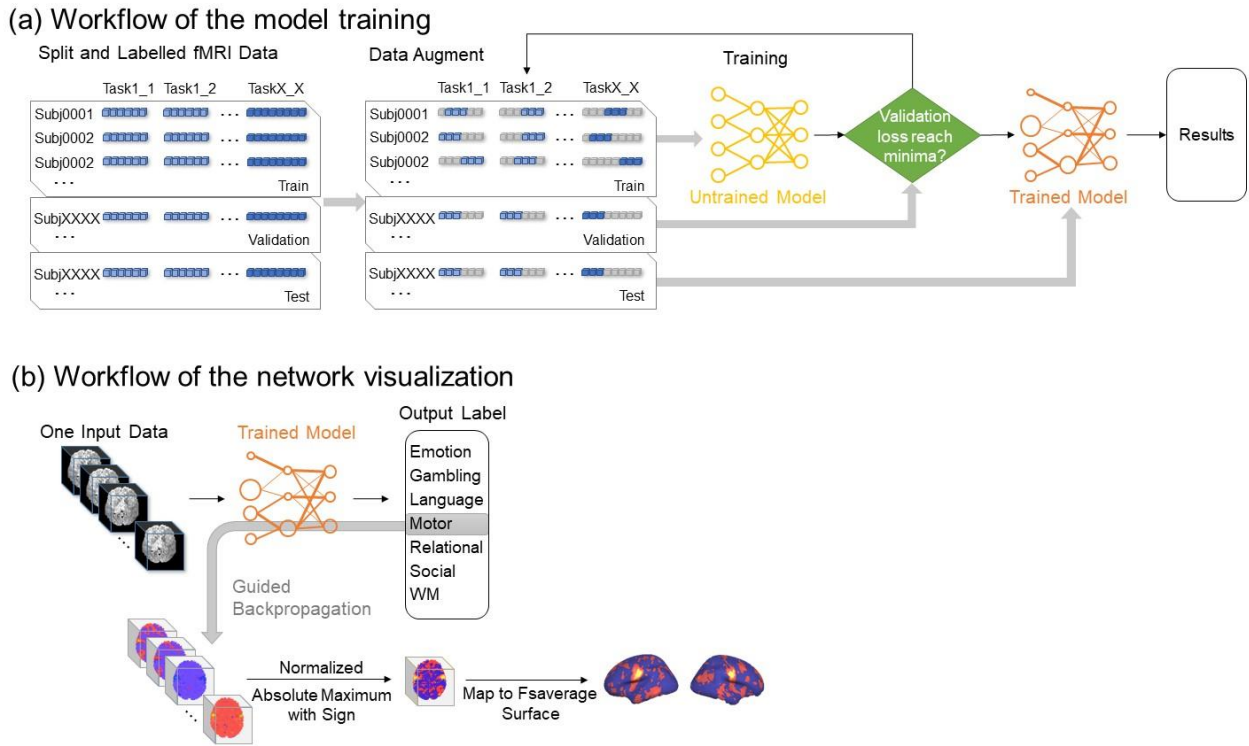

**Figure 2. Workflows of model training and network visualization.** (a) The proposed model automatically learns features of the labeled fMRI time series and stops training when the loss of validation reaches a minimum. Thus, no feature handcrafting is required for model training. The workflow of transfer learning is similar, except that the untrained model is replaced by the trained model. (b) The classification of each data item is back-propagated to the network layers to obtain a visualization of parts important to the classification. The visualized data, which have the same size as the input data, are then reduced in the time dimension and mapped into the fsaverage surface. A motor task data is chosen for the illustration.

## 2.4 Transfer learning

An important advantage of deep learning methods, CNNs in particular, compared with traditional methods, is their reusability, which means that the trained CNN can be directly reused on similar tasks. We used a transfer learning strategy for the trained CNN to validate the general use characteristics of the proposed model. The workflow of transfer training is largely similar to that of the initial training (Figure 2a), except that it starts with a model where the first four layers are trained and the output layer is untrained. We employed the TEST dataset of the TEST-RETEST task-fMRI group from the HCP (N=43). We trained the deep model to classify two WM task sub-states—0bk-body and 2bk-body. A subject-wise five-fold cross validation was applied with 60% (100 samples of 25 subjects) used for training, 20% (36 samples of nine subjects) for validation, and 20% (36 samples of nine subjects) for testing (172 samples in total are comparable in size to commonly used fMRI research datasets). For further validation, we trained the deep model to classify four motor task sub-states—left foot, left hand, right foot, and tongue movement—using five-fold cross validation with 60% (400 samples of 25 subjects) used for training, 20% (144 samples of nine subjects) for validation, and 20% (144 samples of nine subjects) for testing (688 samples altogether). As in the previous scheme, an input sample was a continuous BOLD series that covered the entire block and eight seconds past the block, including the post-signal of the HRF.

For a comparison with the proposed deep learning method, the SVM-MVPA method was also used to analyze the TEST-RETEST dataset using The Decoding Toolbox (Hebart, Gorgen, & Haynes, 2014) in MATLAB (MathWorks, Natick, MA). The run-wise beta images of each subject were obtained through a GLM with separate regressors embedded in the HCP standard FEAT scripts for each task condition. The resulting beta images were then taken as inputs to the SVM-MVPA. A searchlight analysis was also applied: A sphere with a radius of three voxels "searchlight" moved through each brain using a multi-class classification SVM function (fitcecoc, the Statistics and Machine Learning Toolbox of MATLAB) with a linear kernel. The F1 score (see the section "2.6 Assessments") for each condition was calculated as the resulting map. Five-fold cross-validation was also employed. The classifier was trained on data from four-fifths of the subjects and tested on data from the remaining one-fifth.

To evaluate the applicability of the DNN of fMRI studies using small sample sizes, we trained the deep classifiers on data from the 43 subjects of the HCP TEST scans: N=1, 2, 4, 8, 17, 25, 34. To avoid variance in accuracy, all tests were applied to the RETEST data of all 43 subjects in the HCP Test-Retest dataset. The deep learning was stopped after 120 epochs. Searchlight and whole-brain SVM-MVPA methods were also used for comparison.

## 2.5 Performance evaluation

To assess the performance of the model in classifying different tasks, some useful parameters were computed. The F1 score was computed for each task condition as a function of the TP, FP, and FN: $F1 = (2 \times TP)/(2 \times TP + FP + FN, )$. Here, TP is the true positive, FP is the false positive, and FN is the false negative for each label. The receiver operating characteristic (ROC) curve was also calculated for each label by the one-vs-rest approach, with the parameter sensitivity and specificity denoted by: $sensitivity = TP/(TP + FN)$ and $specificity = TN/(TN + FP)$, where TN is the true negative equal to the sum of the TPs of the rest of the labels. **Accuracy was defined as the ratio of the correct predictions to the total number of classifications: $accuracy = (TP + TN)/(TP + FP + TN + FN)$.**

## 2.6 Network visualization analysis

Guided back-propagation (Springenberg, Dosovitskiy, Brox, & Riedmiller, 2014), a widely used deep network visualization method, was applied to produce pattern maps of each classification and task-weighted representation of the input fMRI 4D time series. During standard back-propagation, the partial derivative of a ReLU unit is copied backward if the input to it is positive, and is otherwise set to zero. In guided-back-propagation, the partial derivative of a ReLU unit is copied backward if both the input to it and the partial derivative are positive. Thus, guided back-propagation maintain paths that have a positive influence on the class score and outputs data features that the CNN detects rather than those it does not. As shown in Figure 2b, after feeding data to the trained networks, 27×75×93×81 prediction gradients were produced with respect to the input data. Then, the signed value with an absolute maximum in the time domain for each voxel was drawn out and built up in a 3D task pattern map, which was then normalized to its maximum value. Finally, the pattern map was mapped into the fsaverage surface. In addition, Cohen's d effect for the normalized pattern maps of the test group was calculated as the mean of the pattern maps of each task divided by their standard deviation (st.d.) (Cohen, 1998). Analysis was conducted in AFNI (Cox, 1996), Freesurfer (Fischl, 2012), HCP Connectome Workbench (https://www.humanconnectome.org/software/connectome-workbench), and MATLAB (MathWorks, Natick, MA). For a comparison between the traditional GLM map and the pattern map, we also obtained the Cohen's effect of contrast of parameter estimate (COPE) from the fMRI analysis package of the HCP task.

## 3    Results

### 3.1    The deep model's performance in general task classification

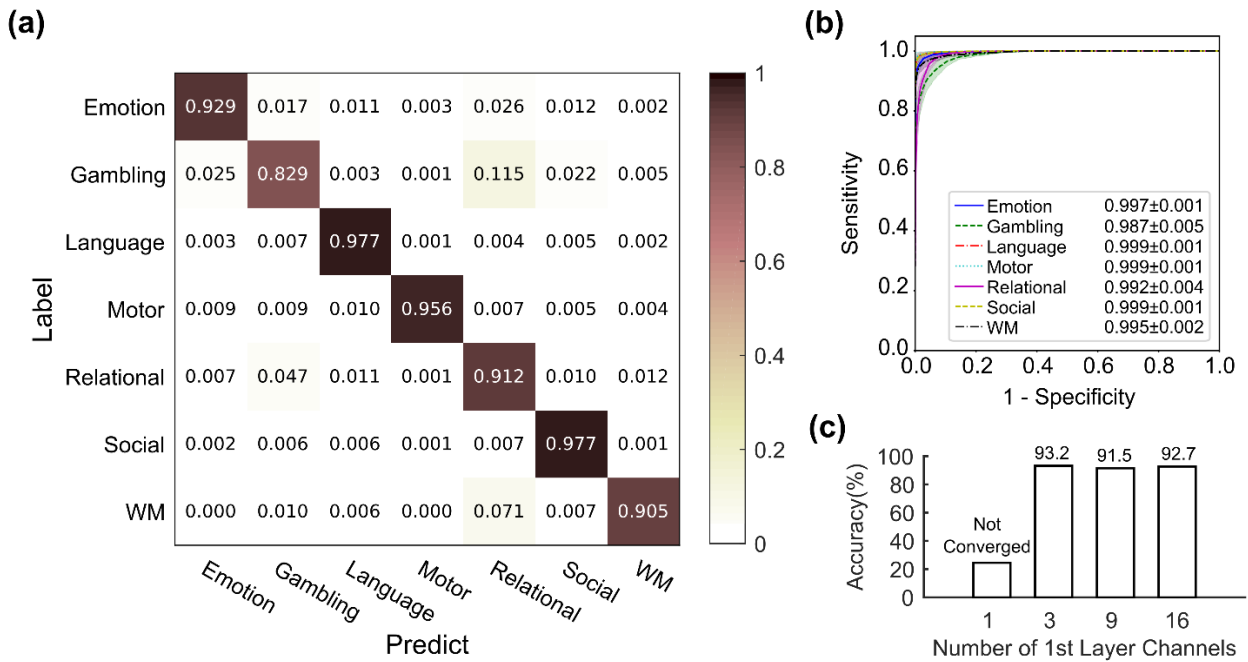

**Figure 3. Results of deep learning classification on the HCP S1200 task fMRI dataset.** (a) The average confusion matrix normalized to the number of labels in the five-fold cross-validation, with the top two

confusions caused by gambling vs. relational and relational vs. WM. The mean (±st.d.) accuracy of classification on the seven tasks was 93.7% (± 1.9%) with a chance level of 14.29%. (b) The mean (solid lines) and st.d. (shadow envelopes) of the ROC curves for each label in the five-fold cross-validation. The legend shows the mean ± st.d. of the AUC of the ROC for the seven tasks. (c) The classification performance (accuracy in %) of the proposed network following various settings of the number of channels in the first layer ($N_{Ch1}$), which was three in the proposed model. The model failed to converge within 30 epochs when $N_{Ch1}=1$.

The training session required approximately 72 hours for the 30 epochs with an NVIDIA GTX 1080Ti board, and the proposed model successfully distinguished seven tasks with an accuracy of 93.7 ± 1.9% (mean ± st.d.). An analysis of F1 scores showed that the classifier performed differently across the seven tasks: emotion (94.0 ± 1.6%), gambling (83.7 ± 4.6%), language (97.6 ± 1.1%), motor (97.3 ± 1.6%), relational (89.8 ± 3.2%), social (96.4 ± 1.0%), and working memory (91.9 ± 2.3%, mean ± st.d.). The average confusion matrix showed that the top two confusions were caused by gambling vs. relational and WM vs. relational (Figure 3a). Figure 3b illustrates the ROC curves, according to which the motor, language, and social task have the largest area under the curve (AUC), while gambling has the smallest.

Upon validation of the choice of key hyper-parameter—the number of 1x1x1 kernel channels ($N_{Ch1}$)––the model recorded accuracy values of 93.2%, 91.5%, and 92.7% with $N_{Ch1}=3$, 9, and 27, respectively (Figure 3c). With $N_{Ch1} =1$, the model could not converge within 30 epochs.

## 3.2 Visualization of learnt patterns

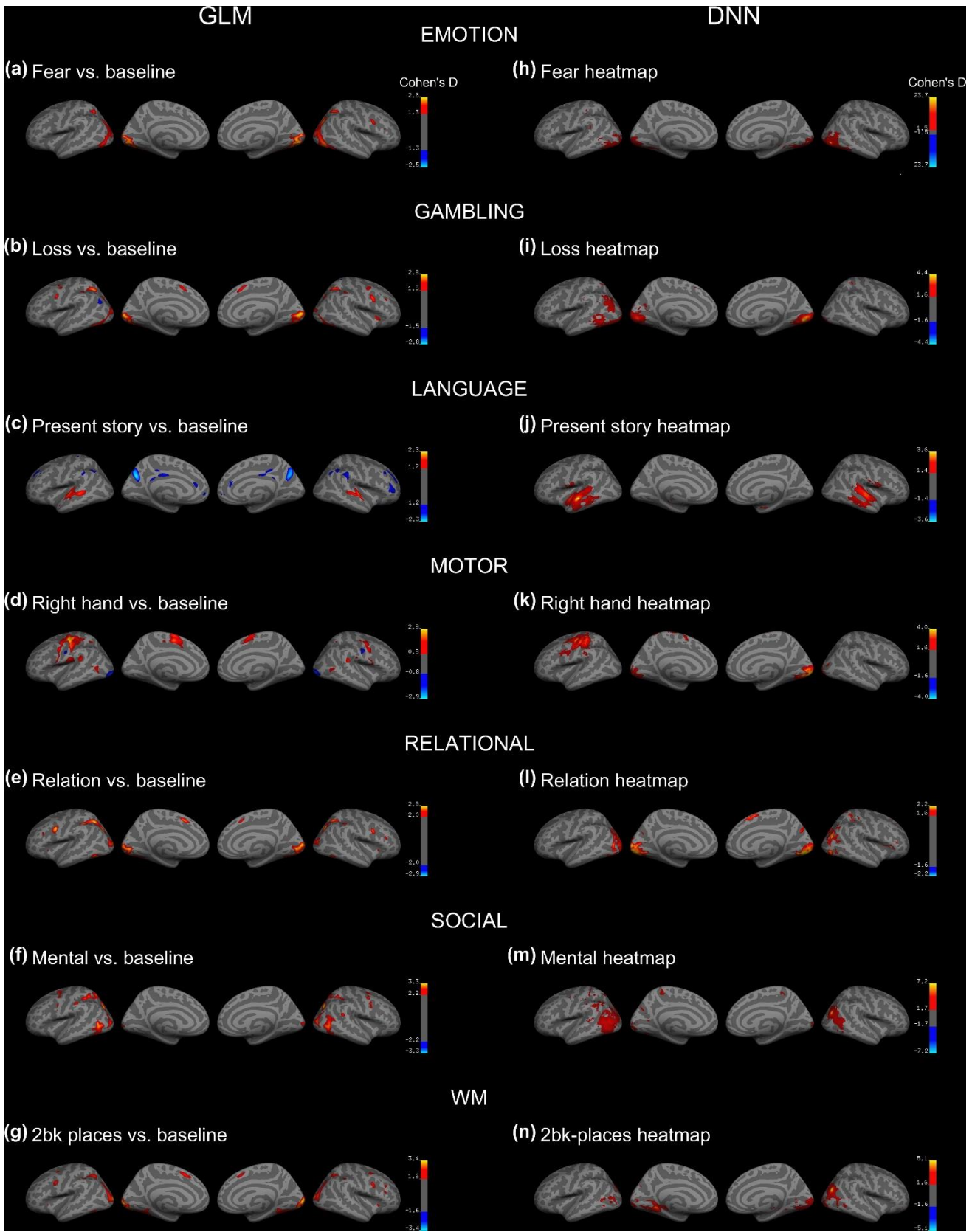

**Figure 4. Cohen's d effect size for the HCP group average (left column) and DNN pattern maps (right column) on the HCP S1200 dataset.**

To identify the voxels contributing most to each classification, we produced pattern maps by using guided back-propagation (Springenberg et al., 2014). Figure 4 shows group statistical maps of the effect size of Cohen's d for the GLM analysis on the task COPE (Figure 4, a–g), and the Cohen's d on the DNN pattern maps (Figure 4, h–n). As shown in the illustrations, the Cohen's d on the DNN pattern maps was similar to that on the GLM COPEs for emotion, language, motor, social, and WM tasks. For example, with the language condition, a large effect size was aberrant in the bilateral Brodmann 22 area in the GLM COPEs (Figure 4c) and DNN pattern maps (Figure 4j). In the same fashion, both maps (Figures 4d and 4k) revealed similar effects in the Brodmann 4 and bilateral Brodmann 18 areas following the right-hand movement condition in the motor task. For further details on annotations, see supplementary Table S1.

.

### 3.3 Transfer learning of WM task sub-types on small datasets

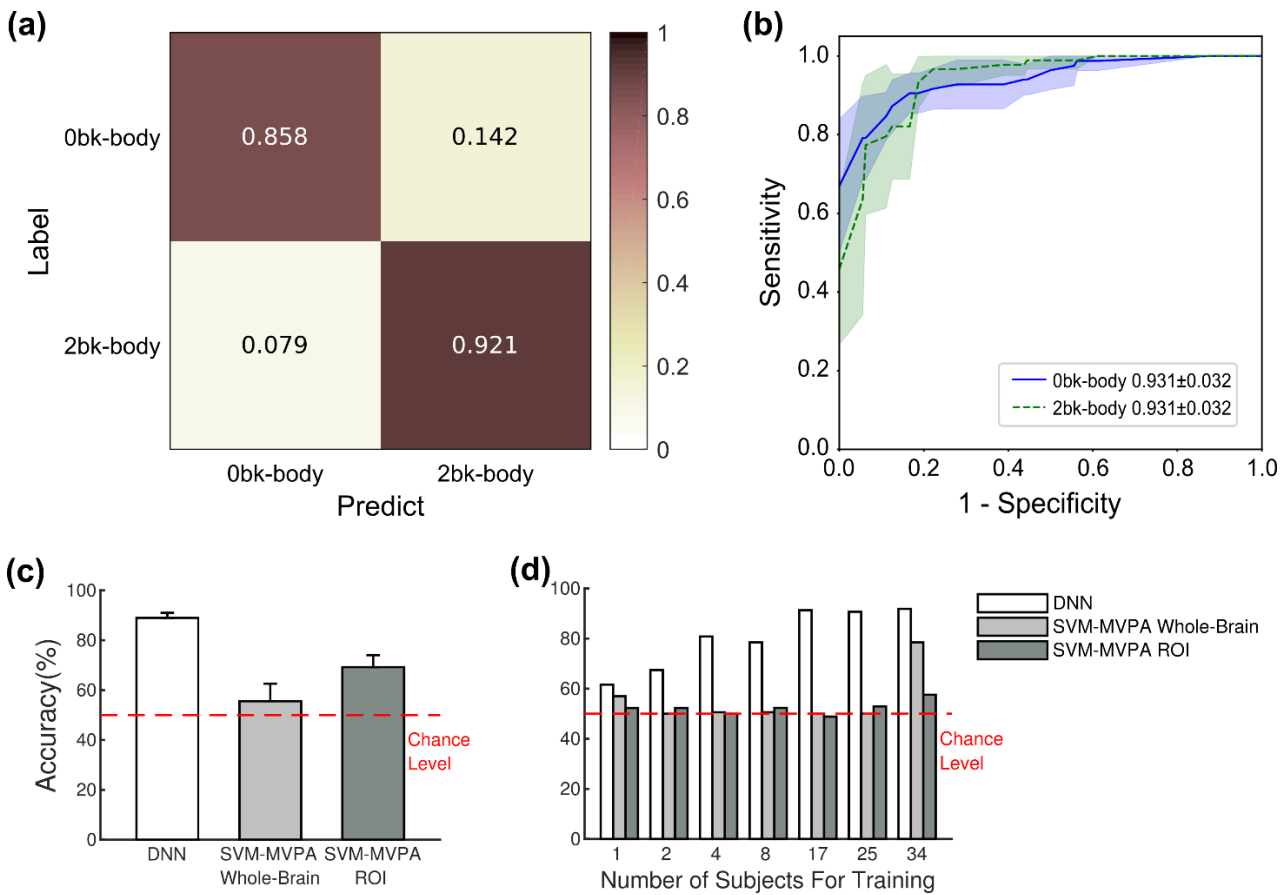

**Figure 5. Results of transfer learning for classification of the working memory task (0bk-body vs. 2bk-body).** (a) The average confusion matrix normalized to the number of instances of each label in five-fold cross-validations. This yielded an average *accuracy* of 89.0 ± 2.0% (mean ± st.d.) in terms of classifying the two tasks (chance level = 50%). (b) The mean (solid lines) and st.d. (shadow envelopes) of the ROC curves for each label in five-fold cross validation. The mean ROC area and st.d. are labeled in the legend. (c) Accuracy of five-fold cross-validation classification on the working memory task on a small dataset. The accuracy of the DNN (89% ± 2%) was significantly higher than that of the SVM-MVPA whole-brain (t(8) = 9.14, p = 0.000017; mean ± st.d. = 55.6 ± 7.9 %) and SVM-MVPA ROI (t(8) = 7.59, p = 0.000064; mean ± st.d. = 69.2 ± 5.4 %) method. (d) The performance of the three methods across different numbers of subjects for training ($N_{Subj}$). $N_{Subj}$ = 2 was enough for the DNN to learn the classification, whereas the SVM-MVPA whole-brain and SVM-MVPA ROI methods needed $N_{Subj}$ = 34.

Following five-fold cross-validation, the proposed DNN reached an average accuracy of 89.0 ± 2.0% (Figure 5a) and an average AUC of ROC 0.931 ± 0.032 (Figure 5b) in the tests. As shown in Figure 5c, the accuracy of the DNN was significantly higher than that of SVM-MVPA whole-brain (t(8) = 9.14, p = 0.000017; mean ± st.d. = 55.6 ± 7.9 %) and SVM-MVPA ROI (t(8) = 7.59, p = 0.000064; mean ± st.d. = 69.2 ± 5.4 %) through a two sample t-test.

We then validated the amount of data needed for learning. The results showed that $N_{Subj}$ = 2 was enough for the DNN to learn the classification (accuracy = 67.4%), whereas SVM-MVPA whole-brain and SVM-MVPA ROI needed $N_{Subj}$ = 34, yielding accuracy values = 91.9%, 78.5%, and 57.6%, respectively (Figure 5d).

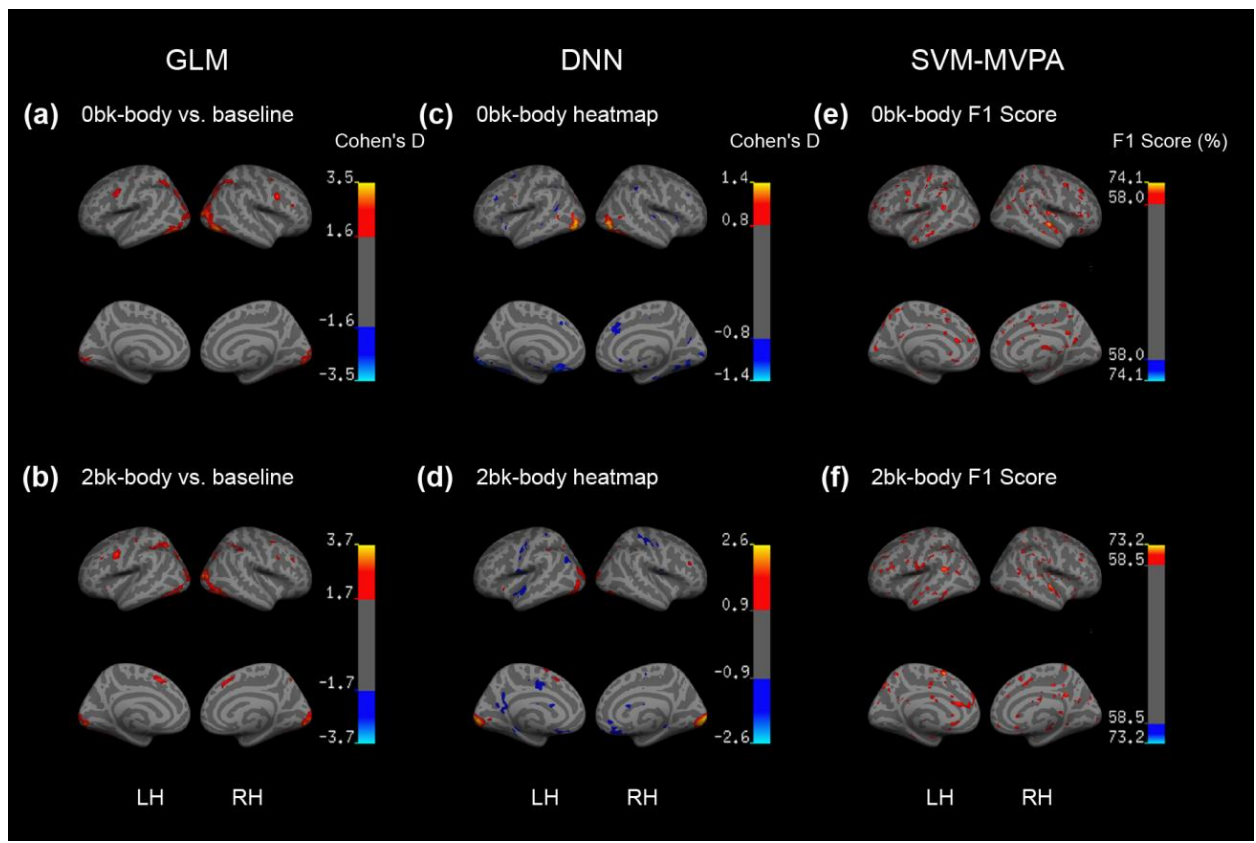

**Figure 6. Visualization of brain task-related maps during the working memory task via GLM, DNN, and SVM-MVPA.** (a, b) Cohen's d for the GLM beta maps. (c, d) Cohen's d for the DNN pattern maps, which showed similar localizations of the fusiform and lateral occipital areas, and dissimilar localizations of lateral and medial orbitofrontal areas, compared with those of the GLM beta maps. (e, f) The F1 score

of the SVM-MVPA searchlight method. It shows that the searchlight failed to localize any functional cluster related to the task but reported widespread scatters all over the brain.

Finally, we visualized the DNN pattern maps and found that the Cohen's d reached its highest value in the Brodmann area 38 (fusiform) and Brodmann area 18/19 (extrastriate visual areas) (Figure 6c, d), which were similar to the results of the GLM COPEs (Figure 6a, b). Moreover, the SVM-MVPA searchlight method reported widespread activity scatters, rather than activity clusters, all over the brain (Figure 6e, f). Refer to supplementary Table S2 for further details on the annotations of the maps.

### 3.4 Transfer learning multiple sub-types of motor task using small datasets

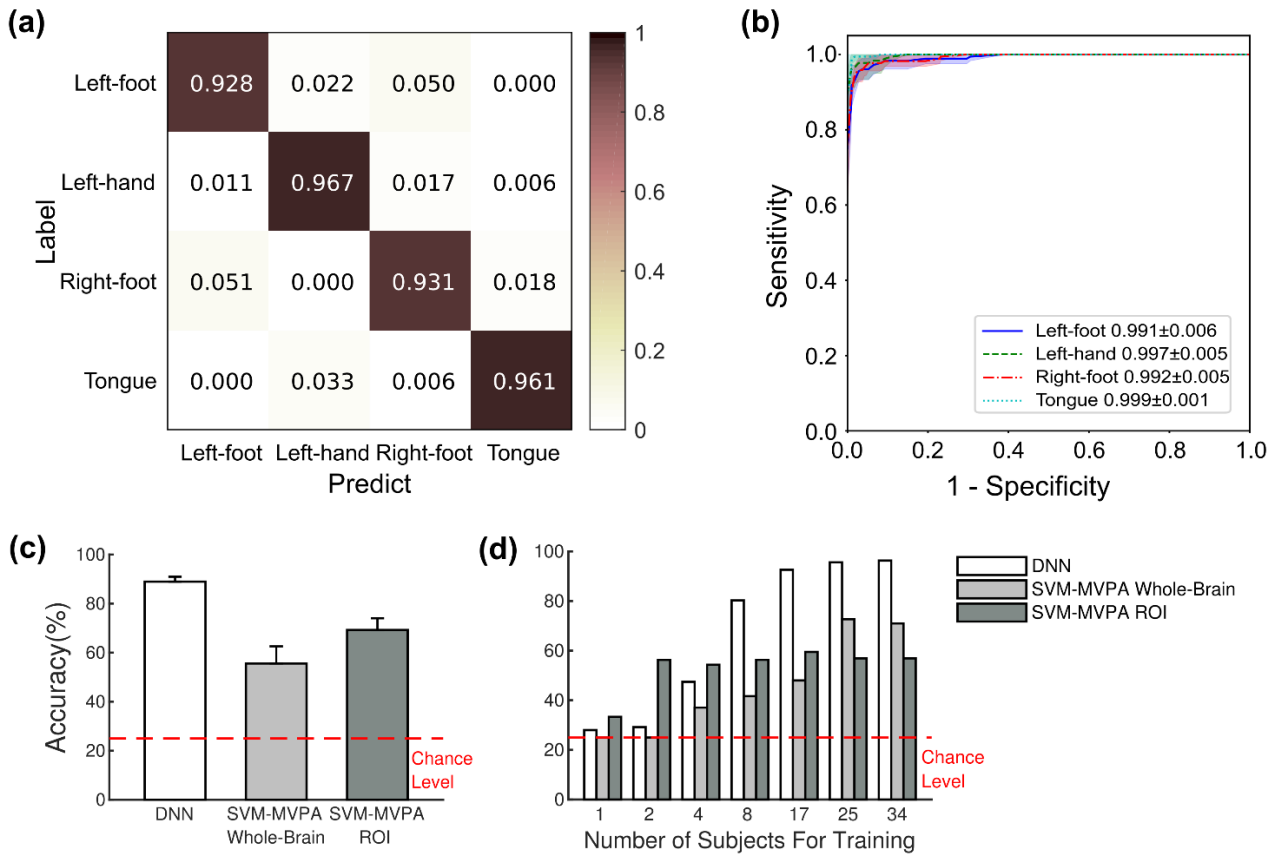

**Figure 7. Results of transfer learning of classification on motor tasks (left foot, left hand, right foot, and tongue).** (a) The average confusion matrices normalized to the number of instances of each label in the five-fold cross-validation, with the top confusion caused by left foot vs. right foot. It reported an average accuracy of 94.7 ± 1.7% (mean ± st.d.) on the four tasks (chance level=25%). (b) The mean (solid lines) and st.d. (shadow envelopes) of ROC curves for each label in the five-fold cross validation. The mean ROC area and st.d. are labeled in the legend. (c) Accuracy of five-fold cross-validation classification on the motor task on a small dataset. The accuracy of the DNN (94.7 ± 1.7%) was significantly higher than that of SVM-MVPA whole-brain (t(8) = 3.59, p = 0.0071; mean ± st.d. = 81.6 ± 7.1%) and SVM-

MVPA ROI (t(8) = 8.77, p = 0.000022; mean ± st.d. = 68.6 ± 5.7%) methods. (d) The performance of the three methods across different numbers of subjects for training ($N_{Subj}$). All conditions reported higher than chance-level accuracy. $N_{Subj}$=8 was enough for the DNN to outperform the ordinary SVM-MVPA methods.

Following five-fold cross-validation, the proposed DNN reached an average accuracy of 94.7 ± 1.7% (Figure 7a) and an average AUC of ROC 0.996 ± 0.005 (Figure 7b). The average confusion matrix showed that the top confusion was caused by left foot versus right foot (Figure 7a). Figure 7c shows that the accuracy of the DNN (94.7 ± 1.7%) was significantly higher than that of SVM-MVPA whole-brain (t(8) = 3.59, p = 0.0071; mean ± st.d. = 81.6 ± 7.1%) and SVM-MVPA ROI (t(8) = 8.77, p = 0.000022; mean ± st.d. = 68.6 ± 5.7%) through a two sample t-test.

We then validated the amount of data needed for learning. All three methods reported higher than chance-level accuracy across all $N_{Subj}$. $N_{Subj}$ = 8 was enough for the DNN (80.3%) to outperform the ordinary SVM-MVPA whole-brain (41.7%) and SVM-MVPA ROI (56.3%) methods in terms of accuracy (Figure 7d).

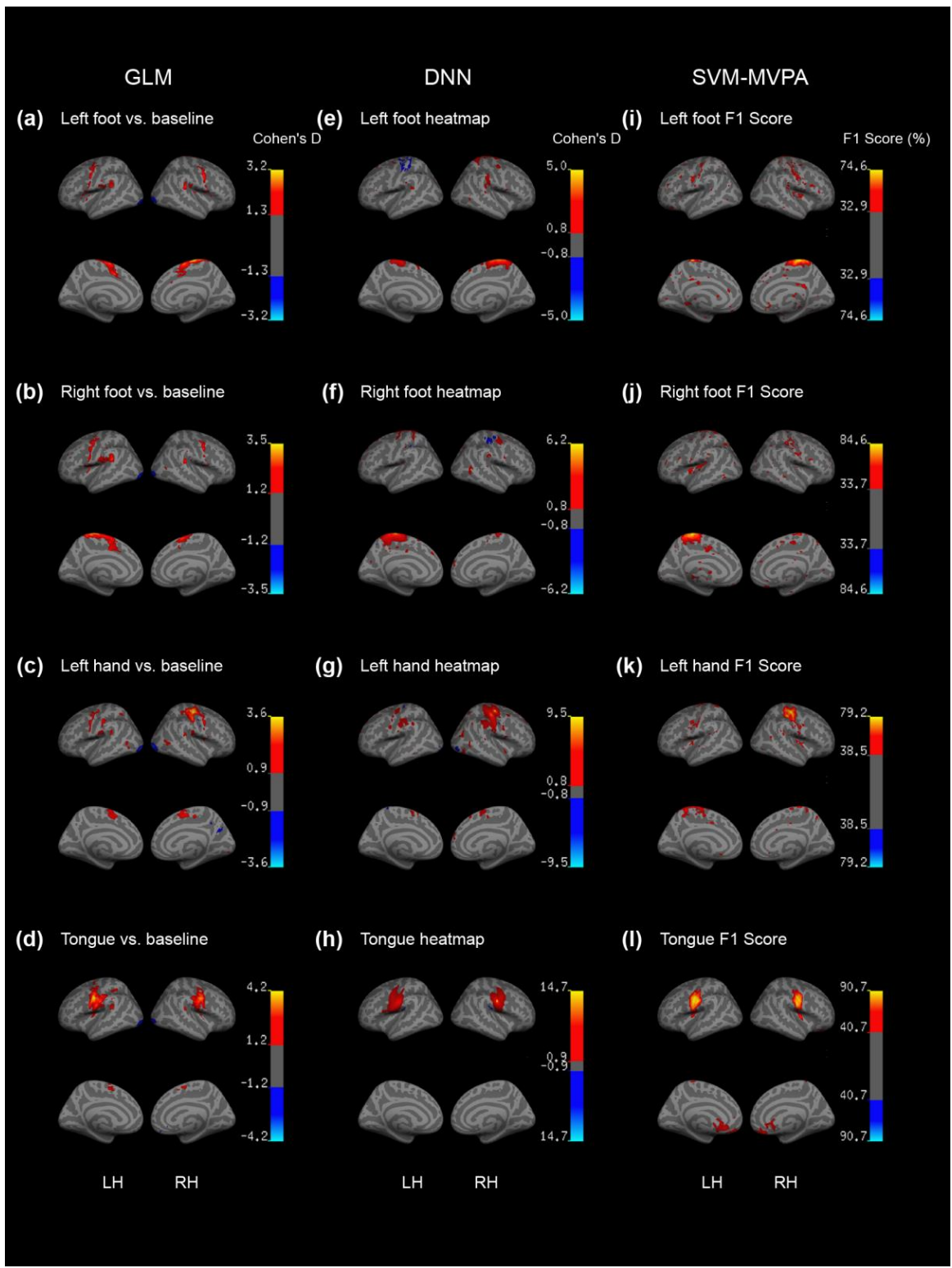

**Figure 8. Visualization of brain task-related maps during motor tasks via GLM, DNN, and SVM-MVPA.** (a–d) Cohen's d effect sizes for the GLM beta maps. (e–h) Cohen's d effect sizes for DNN pattern maps. (i–l) The F1 score of the SVM-MVPA searchlight method. Collectively, the three methods identified similar brain activity maps.

Finally, we visualized the DNN pattern maps and found that Cohen's d reached the highest values in the corresponding motor topological areas (Figures 8e–h), which was similar to the results of the GLM COPEs (Figures 8a–d) and the SVM-MVPA searchlight method (Figures 8i–l). Refer to supplementary Table S2 for further details on the annotations of the maps.

## 4    Discussion

**Summary.** In this study, we proposed a general deep learning framework for decoding and mapping ongoing brain task states from whole-brain fMRI signals of humans. After training and testing it using data from the HCP, the proposed DNN classifier achieved an average accuracy of 93.7% and an average area under the ROC curve of 0.996 on a seven-class classification task. The DNN was able to transfer-learn a new classification task using small fMRI datasets and yielded higher accuracy than SVM-MVPA methods. Moreover, a network visualization analysis showed that the DNN automatically detected and located features in areas of the brain that have been reported to have significant effects in the traditional GLM method.

**Deep learning as a research tool.** Deep learning is capable of automatic data-driven feature learning and has deeper models than earlier methods. Analogous to the brain's sensory network, DNNs perform complex computations through deep stacks of simple intra-layer neural circuits. Thus, researchers have widely used DNN models to understand the human brain network, especially sensory brain networks (Eickenberg, Gramfort, Varoquaux, & Thirion, 2017; Guclu & van Gerven, 2015; Horikawa & Kamitani, 2017; Rajalingham et al., 2018; Yamins & DiCarlo, 2016). At the same time, DNNs are capable of discovering complex structures within high-dimensional input data, and can transform these structures into abstract levels (LeCun et al., 2015). These important features allow researchers to efficiently model complex systems without the burden of model/prior knowledge selection, especially in cases where too many features exist, as when analyzing medical images (Shen et al., 2017). Thus, DNNs are widely used by researchers for medical image analysis, such as brain image segmentation (Havaei et al., 2017; Wachinger et al., 2018; Zhang et al., 2015), neurology and psychiatric diagnostics (Hosseini-Asl, Keynton, & El-Baz, 2016; Meszlenyi, Buza, & Vidnyanszky, 2017; Plis et al., 2014; Vieira et al., 2017), brain state decoding (Jang et al., 2017), and brain computer interfaces (Schirrmeister et al., 2017).

A variety of deep methods have been applied to fMRI data, such as the autoencoder (J. Kim, Calhoun, Shim, & Lee, 2016), deep belief network (DBN) (Jang et al., 2017; Plis et al., 2014), long short-term memory (LSTM) recurrent neural network (RNN) (Li & Fan, 2019), and 2D CNN (Meszlenyi et al., 2017). Although the autoencoder is known to be efficient, especially when the dataset is small, it over-emphasizes some relationships while neglecting others, i.e., it loses information. DBNs have been criticized for a number of shortcomings, such as the computational cost associated with training and loss of spatial information in learning, which may significantly affect their performance and interpretability in medical image analysis (Voulodimos, Doulamis, Doulamis, & Protopapadakis, 2018). The RNN with LSTM, a deep learning method for sequence modeling, ignores spatial information within the input data (Hochreiter & Schmidhuber, 1997). The 2D CNN cannot encode the 3D nature of fMRI data. Thus, both Li and Fan (2019) and Meszlenyi et al. (2017) methods require functional network-based features as inputs. Our study represents a significant departure from these studies, however, by directly targeting fMRI volume through the 3D CNN. The proposed 3D CNN, which makes use of the spatial structure of the input data, is efficient in capturing spatial relationships of the brain activity. As end-to-end learning methods, CNNs have the unique capability of learning features automatically and avoids the design of a feature extractor. On the contrary, CNNs heavily rely on manually labeled training data, but this is not a problem for neuroimaging research because almost all neuroimaging data are carefully labeled with diagnostics, task states, and questionnaires. Moreover, because the CNN requires scant handcrafting of features by experts, it is easily usable by data scientists on neuroimaging data.

We used an NVIDIA GTX 1080Ti GPU in our experiments. The initial training took a long time (72 hours for 30 epochs) while transfer learning took much less time (9 hours for 120 epochs on the two-class classification task, and 21 hours for 120 epochs in the four-class classification task). The proposed CNN was composed of three convolutional layers and two fully connected layers with 3,981,852 parameters. Given these layers and their hyperparameters, we could make countless possible combinations of network architectures. We evaluated the impact of the number of 1x1x1 channels (Figure 3c), and found that three channels provided enough information to distinguish between task states. The proposed model was implemented on the PyTorch library: a free and open-source software and among the most popular deep learning platforms. Researchers interested in reusing the proposed model on other platforms can refer to the Open Neural Network Exchange (ONNX) created by Facebook and Microsoft.

**Visualization of learnt patterns.** The proposed method also offers researchers the opportunity to investigate decisions of the neural network. A challenge of applying deep models to neuroimaging research is the black-box characteristic of this approach: No one knows exactly what the deep network is doing. In recent years, a method for tracing consecutive layers of weights back to the original image inputs has been proposed, and has achieved good performance in natural image recognition (Springenberg et al., 2014). Researchers have employed various methods for the analysis of the processes of deep neural networks (Bach et al., 2015; Yamins & DiCarlo, 2016). Guclu and van Gerven (2015) employed a DNN model to predict the responses of each voxel and found a gradient in the feature complexity aligning with the ventral pathway. Through linear predictive models, Eickenberg et al. (2017) generalized human visual cortical activity maps elicited by visual stimulation. Jang et al. (2017) proposed a ROI-wise task-specific activity map by extracting the weights of the nodes in the output layer of a deep network.

We employed guided back-propagation, a widely used network visualization method, to visualize features of the data detected by the CNN for the classification of each entered data item. The visualized voxels with values other than zero comprised features important for classification. There is a criticism where good decoding performance is not a guarantee that patterns of brain activity are learned (Ritchie, Kaplan, & Klein, 2019), for a decoder may learn from nuisance or latent variables (P. Riley, 2019)—for example, the different visual responses to different stimulus images or patterns of response key-pressing across the seven tasks. The guided back-propagation allows scientists to intuitively locate and investigate features the DNN detected in every entered fMRI data item. In this work, the similarity between the pattern maps and the GLM maps (Figures 4, 6, and 8) suggest that the proposed DNN decoded states from task-related brain activity patterns, not from nuisance variables. Furthermore, correlated with the β maps of the GLM, the pattern maps showed potential for localizing state-related areas of the brain. However, the statistical property of guided back-propagation remains unclear, and we should be cautious until further investigations on its reliability and statistical properties.

**Transfer learning helps model construction with small samples.** Transfer learning is a machine learning method that learns from networks trained on a related but different task from the given one. By taking advantage of transferred knowledge, it eliminates the need for big training data (Rawat & Wang, 2017). Hosseini-Asl et al. (2018) pre-trained a 3D convolutional autoencoder to capture anatomical shape variations in brain MRI scans and fine-tuned it for AD classification on images from 210 subjects. Gao et al. (2019) pre-trained a 2D-CNN for classification on ImageNet, a database containing > 14 million natural images, and fine-tuned it to decode 2D fMRI slices. The proposed method transfer-learns in a more direct way—transferring knowledge learnt from a big fMRI dataset to limited fMRI datasets. We believe that the proposed DNN can transfer-learn a related but different decoding task using fMRI data from as few as four subjects (Figure 5d). Although our deep learning framework was trained and validated using the HCP S1200 dataset, the consistent internal properties of human haemodynamic responses make fMRI data reasonably consistent across scanners and sites. Nowadays big datasets, such as BioBank, HCP, and OpenfMRI, provide comprehensive neuroimaging scans across a wide range of ages and diseases, and provide the opportunity for pretraining on big data and transfer learning on small fMRI datasets.

**Transfer learning to the working memory task.** We evaluated the generalizability of our deep learning framework in transfer learning to working memory data of 43 subjects. Working memory refers to a brain function for the temporary storage and manipulation of information for cognitive processing (Baddeley, 1992). We chose the working memory because researches have shown that it is not processed in a single brain site, but stored and processed in widely distributed brain regions (Christophel, Klink, Spitzer, Roelfsema, & Haynes, 2017; Mencarelli et al., 2019), ranging from the sensory (Pasternak, Lui, & Spinelli, 2015; Sreenivasan, Curtis, & D'Esposito, 2014) to prefrontal (Durstewitz, Seamans, & Sejnowski, 2000; M. R. Riley & Constantinidis, 2015) and parietal (Xu & Jeong, 2016) cortices. This distributed nature of the working memory makes it impossible to decode from a single ROI, as shown in this work, and poses a major obstacle to ROI selection in the MVPA. We proposed a machine learning framework that automatically abstracted the activity patterns of the brain, affording a powerful tool to decode comprehensive brain functions. Moreover, by using guided back-propagation, we showed that the proposed model detected features from areas of the brain that have been reported to be related to the working memory function: BA 32 (anterior cingulate cortex, Owen, McMillan, Laird, and Bullmore (2005)), BA 38 (fusiform, Downing, Jiang, Shuman, and Kanwisher (2001); Kanwisher, McDermott, and Chun (1997)), and BA 18/19 (extrastriate visual cortex, Grill-Spector, Kourtzi, and Kanwisher (2001)). Its performance in classifying two tasks provided more evidence that the model learnt from task-related brain activity, rather than nuisance variables, because the stimuli were consistent, with merely the task altered, between 0-back and 2-back.

**Transfer learning to the motor task.** We evaluated the generalizability of our deep learning framework in transferring learning to multi-class motor data of 43 subjects. Motor-related information was encoded in the primary motor cortex, premotor cortex, and supplementary motor area around the central sulcus. The topological nature of the motor area makes it the first cortex to be decoded in the human brain (Dehaene et al., 1998). In our experiment, the SVM-MVPA was good at single-label classification (high F1 scores for each task in Figure 8) but delivered poor performance at multi-class classification (low accuracy in Figure 7d). The proposed method showed its potential in multi-class classification over the SVM-MVPA method. Cognitive neuroscience has attended to particular brain functions, but researchers are now calling for models that generalize beyond specific tasks (Varoquaux & Poldrack, 2019; Yarkoni & Westfall, 2017). Brain systems are often engaged in a variety of brain functions (Varoquaux et al., 2018), and predictive investigations of general tasks can ultimately lead to a greater understanding of the human brain. The proposed method provides researchers with the choice of decoding and interpreting brain functions in an integrative way.

**Future work.** Although we illustrated the deep model's ability to read the fMRI time series, researchers can modify the input layer and take a volume of brain features as input to the proposed deep model, such as the amplitude of low-frequency fluctuation (ALFF), fractional ALFF (fALFF), and regional homogeneity (ReHo) of resting-state fMRI as well as the fractional anisotropy (FA) and mean diffusivity (MD) of diffusion tensor imaging (DTI). The model is also applicable to multi-modal inputs to different channels, which are important for research in psychiatry and neurology because most of the open datasets used, such as ADNI (Alzheimer's Disease Neuroimaging Initiative), ABIDE (Autism Brain Imaging Data Exchange), BioBank, and SchizConnect. The proposed method can provide a basis for a brain-based information retrieval systems by classifying brain activity into different categories: brain-based disorder or psychiatric classification. Varieties of deep learning methods have shown their power in searching for biomarkers of psychiatric and neurologic diseases (Vieira et al., 2017), and the proposed method provides one more choice.

Activity classification can also benefit real-time fMRI neurofeedback (rt-fMRI-NF), a technology providing subjects with feedback stimuli from ongoing brain activity collected by an MRI scanner (Cox, Jesmanowicz, & Hyde, 1995; Sulzer et al., 2013). Recently, a data-driven and personalized MVPA rt-fMRI-NF method (Shibata, Watanabe, Sasaki, & Kawato, 2011), decoded neurofeedback (DecNef), was proposed, and has shown outstanding performance in both basic and clinical research (Thibault, MacPherson, Lifshitz, Roth, & Raz, 2018; Watanabe, Sasaki, Shibata, & Kawato, 2017). The proposed deep model has the potential to decode multiple brain states from whole-brain fMRI time series and to output these to feedback processing in real time. Moreover, the model can be fine-tuned to individual brain activity through transfer learning to build up a personalized rt-fMRI-NF.

**Conclusion.** We proposed a method to classify and map an individual's ongoing brain function directly from a 4D fMRI time series. Our approach allows for the decoding of a subject's task state from a short fMRI scan without the burden of feature selection. This flexible and efficient brain-decoding method can be applied to both large-scale massive data and fine, small-scale data in neuroscience. Moreover, its characteristics of facility, accuracy, and generalizability allow the deep framework to be easily applied to a new population as well as a wide range of neuroimaging research, including internal mental state classification, psychiatric disease diagnosis, and real-time fMRI neurofeedback.

## 5    Data Availability Statement

All scripts described in this paper are available at https://github.com/ustc-bmec/Whole-Brain-Conv.

## 6    Conflict of Interest Statement

The authors declare that the research reported here was conducted in the absence of any commercial or financial relationships that can be construed as potential conflicts of interest.

## 7    Contributions

XW and XL analyzed the data and wrote the paper. ZJ, BAN, YZhou, YW, HW, YL, Yzhu, and FW processed and analyzed the data. JG and BQ conceived of the study and contributed to writing the manuscript. All authors discussed the results and reviewed the manuscript.

## Supplementary

*S-Table 1, localization of heatmaps for both GLM and DNN.*

| Task | GLM | | | | | | DNN | | | | | |
|---|---|---|---|---|---|---|---|---|---|---|---|---|
| | Annotation | TalX | TalY | TalZ | Size (mm²) | Max Cohen's d | Annotation | TalX | TalY | TalZ | Size (mm²) | Max Cohen's d |
| **Emotion** | r BA.37 | 39 | -58 | -13 | 6120 | 2.49 | r BA.19 | 43 | -79 | -3 | 5383 | 23.69 |
| | l BA.19 | -36 | -71 | -10 | 4886 | 2.39 | l BA.19 | -35 | -77 | -7 | 4487 | 2.47 |
| **Gambling** | l BA.17 | -14 | -89 | 7 | 3718 | 2.79 | l BA.18 | -4 | -85 | 0 | 6380 | 3.05 |
| | r BA.17 | 14 | -85 | 10 | 3624 | 2.81 | r BA.18 | 12 | -81 | -3 | 3607 | 4.37 |
| | r BA.7 | 29 | -53 | 44 | 454 | 2.24 | l BA.7 | -37 | -67 | 44 | 1732 | 1.98 |
| | l BA.7 | -26 | -53 | 43 | 424 | 2.49 | | | | | | |
| **Language** | l BA.22 | -52 | -19 | 4 | 1372 | 1.82 | l BA.22 | -52 | -16 | -6 | 4345 | 3.60 |
| | r BA.22 | 56 | -13 | 2 | 835 | 1.80 | r BA.22 | 60 | -14 | 3 | 3327 | 2.74 |
| | r BA.31 | 11 | -61 | 32 | 700 | -2.14 | r BA.6 | 48 | 0 | 10 | 440 | 1.64 |
| | l BA.19 | -12 | -62 | 33 | 673 | -2.30 | l BA.43 | -60 | -10 | 11 | 422 | 1.88 |
| | r BA.46 | 37 | 38 | 5 | 425 | -1.52 | | | | | | |
| **Motor** | l BA.4 | -36 | -16 | 51 | 3170 | 2.86 | l BA.4 | -37 | -14 | 52 | 3801 | 3.50 |
| | l BA.18 | -28 | -92 | -5 | 1072 | -1.88 | r BA.18 | 10 | -88 | -2 | 3758 | 4.04 |
| | r BA.18 | 29 | -91 | -1 | 880 | -1.84 | l BA.19 | -19 | -78 | -2 | 1740 | 2.07 |
| | l BA.6 | -6 | -3 | 50 | 829 | 2.05 | | | | | | |
| | r BA.6 | 45 | -1 | 42 | 523 | 1.50 | | | | | | |
| | r BA.6 | 7 | 2 | 60 | 489 | 1.50 | | | | | | |
| | l BA.43 | -54 | -18 | 21 | 480 | 1.42 | | | | | | |
| **Relational** | l BA.17 | -5 | -82 | 3 | 3134 | 2.83 | l BA.18 | -12 | -86 | -5 | 6763 | 2.20 |
| | r BA.18 | 5 | -82 | 2 | 2200 | 2.86 | r BA.18 | 15 | -85 | -6 | 3144 | 2.23 |
| | l BA.7 | -26 | -54 | 43 | 1360 | 2.93 | r BA.19 | 43 | -75 | 21 | 1098 | 2.17 |
| | r BA.19 | 31 | -61 | 28 | 815 | 2.81 | r BA.18 | 26 | -71 | 26 | 779 | 1.95 |
| | r BA.18 | 15 | -94 | 16 | 737 | 2.43 | | | | | | |
| **Social** | l BA.19 | -28 | -66 | 23 | 1401 | 3.19 | l BA.39 | -44 | -68 | 10 | 4982 | 2.80 |
| | r BA.19 | 40 | -64 | 3 | 1173 | 2.90 | r BA.19 | 39 | -78 | 24 | 2324 | 7.21 |
| | l BA.19 | -42 | -63 | 2 | 927 | 3.04 | | | | | | |
| | r BA.7 | 31 | -46 | 48 | 708 | 2.65 | | | | | | |
| | r BA.19 | 29 | -80 | 7 | 691 | 2.82 | | | | | | |
| | l BA.18 | -26 | -84 | 7 | 626 | 2.67 | | | | | | |
| | r BA.19 | 30 | -63 | 25 | 506 | 3.35 | | | | | | |
| **WM** | r BA.17 | 16 | -89 | 4 | 4317 | 3.42 | r BA.19 | 30 | -77 | -3 | 3108 | 2.49 |
| | l BA.17 | -15 | -94 | 4 | 4155 | 3.30 | r BA.19 | 39 | -77 | 25 | 1978 | 5.14 |
| | | | | | | | l BA.19 | -19 | -49 | -4 | 1533 | 3.88 |
| | | | | | | | l BA.18 | -8 | -92 | -3 | 413 | 1.85 |

S-Table 2. Annotation of in transfer learning to Working Memory

| Task | | GLM | | | | | | | CNN | | | | | | | SVM-MVPA | | | | | |
|---|---|---|---|---|---|---|---|---|---|---|---|---|---|---|---|---|---|---|---|---|---|
| | | Annotation | TalX | TalY | TalZ | Size (mm²) | Max Cohen's d | | Annotation | TalX | TalY | TalZ | Size (mm²) | Max Cohen's d | | Annotation | TalX | TalY | TalZ | Size (mm²) | Max F1 Score |
| **0bk body** | r | BA.37 | 45 | -70 | -5 | 5343 | 3.47 | r | BA.19 | 43 | -77 | 1 | 1243 | 1.40 | r | BA.22 | 62 | -7 | 0 | 447 | 74.11 |
| | l | BA.37 | -42 | -68 | -11 | 2110 | 3.34 | l | BA.19 | -41 | -79 | 0 | 805 | 1.37 | | | | | | | |
| | l | BA.17 | -10 | -99 | 8 | 1125 | 2.66 | l | BA.32 | -6 | 20 | -18 | 776 | -1.24 | | | | | | | |
| | | | | | | | | l | BA.18 | -17 | -84 | -5 | 731 | -1.23 | | | | | | | |
| | | | | | | | | l | BA.37 | -28 | -54 | -9 | 512 | -1.10 | | | | | | | |
| **2bk body** | r | BA.19 | 27 | -81 | -4 | 3893 | 3.57 | l | BA.18 | -10 | -91 | 0 | 3482 | 2.46 | l | BA.9 | -7 | 53 | 27 | 418 | 65.11 |
| | l | BA.19 | -26 | -80 | -4 | 3516 | 3.71 | r | BA.18 | 14 | -89 | 1 | 2558 | 2.61 | | | | | | | |
| | l | BA.7 | -26 | -54 | 42 | 604 | 2.77 | r | BA.3 | 42 | -19 | 46 | 523 | -1.11 | | | | | | | |
| | | | | | | | | l | BA.22 | -57 | -5 | -5 | 406 | -1.36 | | | | | | | |

S-Table 3. Annotation of in transfer learning to Motor

| Task | GLM | | | | | | CNN | | | | | | SVM-MVPA | | | | | |
|---|---|---|---|---|---|---|---|---|---|---|---|---|---|---|---|---|---|---|
| | Annotation | TalX | TalY | TalZ | Size (mm²) | Max Cohen's d | Annotation | TalX | TalY | TalZ | Size (mm²) | Max Cohen's d | Annotation | TalX | TalY | TalZ | Size (mm²) | Max F1 Score |
| left foot | r BA.4 | 6 | -19 | 68 | 1920 | 3.18 | r BA.4 | 6 | -26 | 67 | 3010 | 4.99 | l BA.4 | 3 | -31 | 65 | 1838 | 74.61 |
| | l BA.6 | -9 | 0 | 58 | 960 | 2.41 | l BA.1 | -46 | -16 | 48 | 1046 | -1.34 | l BA.1 | 51 | -12 | 33 | 793 | 43.37 |
| | r BA.18 | 29 | -92 | -3 | 676 | -1.98 | l BA.4 | -6 | -12 | 63 | 781 | 1.89 | | | | | | |
| | l BA.6 | -54 | -1 | 36 | 624 | 1.99 | r BA.42 | 47 | -36 | 22 | 589 | 1.04 | | | | | | |
| | r BA.6 | 55 | 1 | 35 | 508 | 2.14 | | | | | | | | | | | | |
| | l BA.18 | -29 | -92 | -6 | 488 | -2.17 | | | | | | | | | | | | |
| left hand | r BA.4 | 33 | -18 | 44 | 2669 | 3.59 | r BA.4 | 38 | -14 | 54 | 4631 | 9.48 | r BA.3 | 36 | -23 | 44 | 2526 | 79.23 |
| | r BA.18 | 28 | -92 | -1 | 976 | -2.37 | l BA.2 | -47 | -23 | 37 | 714 | 1.04 | l BA.4 | -4 | -30 | 65 | 1870 | 54.25 |
| | l BA.18 | -30 | -91 | -4 | 830 | -2.36 | l BA.4 | -39 | -13 | 57 | 467 | 1.36 | | | | | | |
| | l BA.4 | -40 | -10 | 55 | 756 | 1.79 | r BA.19 | 36 | -73 | -7 | 401 | 1.74 | | | | | | |
| | l BA.6 | -8 | 1 | 59 | 511 | 1.72 | | | | | | | | | | | | |
| | r BA.6 | 7 | -4 | 52 | 463 | 1.72 | | | | | | | | | | | | |
| right foot | l BA.3 | -5 | -33 | 64 | 2224 | 3.52 | l BA.4 | -8 | -26 | 66 | 3387 | 6.21 | l BA.4 | -3 | -33 | 63 | 2282 | 84.58 |
| | l BA.40 | -48 | -42 | 24 | 1059 | 2.40 | r BA.1 | 42 | -22 | 57 | 491 | -1.35 | l BA.41 | -31 | -25 | 17 | 482 | 44.26 |
| | l BA.4 | -43 | -5 | 42 | 905 | 2.08 | r BA.4 | 40 | -8 | 49 | 438 | 1.37 | | | | | | |
| | l BA.18 | -29 | -92 | -6 | 620 | -2.31 | | | | | | | | | | | | |
| | r BA.18 | 24 | -96 | -6 | 602 | -2.03 | | | | | | | | | | | | |
| | r BA.6 | 7 | 4 | 60 | 594 | 2.33 | | | | | | | | | | | | |
| tongue | l BA.4 | -49 | -5 | 30 | 3179 | 4.24 | l BA.1 | -57 | -8 | 27 | 3641 | 4.52 | l BA.3 | -46 | -7 | 24 | 2437 | 90.72 |
| | r BA.4 | 49 | -4 | 28 | 2308 | 3.43 | r BA.1 | 58 | -8 | 29 | 2860 | 14.74 | r BA.3 | 47 | -8 | 24 | 1944 | 86.20 |
| | l BA.18 | -24 | -94 | -3 | 603 | -2.30 | | | | | | | r BA.11 | 5 | 34 | -21 | 755 | 59.09 |
| | r BA.18 | 20 | -97 | -1 | 575 | -2.25 | | | | | | | l BA.11 | -4 | 43 | -20 | 625 | 53.42 |
| | | | | | | | | | | | | | l BA.24 | -8 | 26 | -12 | 480 | 50.81 |